\documentclass[12pt,oneside]{amsart}
\usepackage{euscript,mathtools,amsthm,amsfonts,amssymb}
\usepackage[curve,knot]{xy}
\usepackage{subcaption}
\usepackage{cite}

\numberwithin{equation}{section}

\begin{document}
\title{Interferometric versus projective measurement of anyons}
\author{Michael Freedman \and Claire Levaillant}

\begin{abstract}
Two distinct methods for measuring topological charge in a nonabelian anyonic system have been discussed in the literature: projective measurement of a single point-like quasiparticle and interferometric measurement of the total topological charge of a group of quasiparticles.  Projective measurement by definition is only applied near a point and will project to a topological charge sector near that point.  Thus, if it is to be applied to a \emph{group} of anyons to project to a \emph{total} charge, then the anyons must first be fused one by one to obtain a single anyon carrying the collective charge.  We show that interferometric measurement is strictly stronger: Any protocol involving projective measurement can be simulated at low overhead by another protocol involving only interferometric measurement.
\end{abstract}

\maketitle

\section{}

We clarify a foundational issue regarding the relative power of two computational schemes which have been long proposed for nonabelian anyons.  The two are projective measurement \cite{K} and interferometric measurement; abstractly \cite{P,OB,B3,B2,B1} and in a physical system, \cite{DFN}.

The discussion is restricted to the basic case: unitary modular tensor categories (UMTC) \cite{W}.  This is a mathematical idealization of a ($2+1$)-dimension topological quantum field theory (TQFT).  The unitary requirement is necessary for the system to be a legitimate model of the non-dissipative physics of the low energy states of a gapped quantum mechanical system.  The modularity assumption is a nonsingularity requirement.  Abstractly this condition assures us that the quantum system can be consistently defined on a physical torus (and this implies all surfaces \cite{T})---the system is \emph{not} tied to the plane.  Concretely it tells us that every topologically nontrivial excitation (or equivalently ``quasiparticle'' or ``anyon'') of the theory $a$ is detected by its braiding with some other anyon $b$ (see page 29 of \cite{B1}).  For example, when anyons $a$ and $b$ are mutually abelian, i.e., have a unique fusion channel $c$, braiding assumes the form:
\[\begin{xy}<5mm,0mm>:
(0,0)*{\vtwist}
,(0,0.75)*{\vtwist}
,(0,-1.5)*{b}
,(1,-1.61)*{a}
\ar (0,0);(0,0.5)
\ar (1,0);(1,0.5)
\end{xy}\; = M_{b,a}\;
\begin{xy}<5mm,0mm>:
(0,-1.5)*{b}
,(1,-1.61)*{a}
\ar (0,-1);(0,0.5)
\ar@{-} (0,0.5);(0,1.5)
\ar (1,-1);(1,0.5)
\ar@{-} (1,0.5);(1,1.5)
\end{xy}\text{, $M_{b,a}$ a nontrivial phase.}\]

Modularity implies that for each nontrivial particle $a$, there is a (possibly composite) $b$ so that not only is $M_{b,a}\neq 1$ but $M_{b,a}\neq M_{b,a^\prime}$, $a^\prime = a$, $a^\prime$ an anyon of the theory.  Even for \emph{non-abelian} theories, modularity implies that particles are distinguished via braiding.

This distinguishability was exploited in \cite{B1,B2} to give a full analysis of anyonic interferometers of Mach-Zehnder and Fabry-Perot designs.  We note that much of the applied and even experimental literature on measuring topological charge, e.g. \cite{Wi}, has focused on Fabry-Perot geometries.  In the low tunnelling (single pass) limit, the Fabry-Perot interferometer determines the same evolution of the system density matrix $\rho$ as the simpler Mach-Zehnder interferometer.  Even so, the evolution of $\rho$ under operation of the interferometer is quite complicated.  Fortunately, and this is the main conclusion of \cite{B1,B2}, there is an easy \emph{topological} interpretation of the asymptotic action of an interferometer, $\operatorname{Int}_a$, which has converged to a measurement ``a'', meaning the total topological charge within the interferometric loop has been measured to be that of an anyon of type $a$.  See Figure \ref{fig:1} below.  Furthermore, convergence to this limit is efficient---exponentially fast.

\begin{figure}[hbpt]
\begin{subfigure}{\textwidth}
\[\begin{xy}<5mm,0mm>:
(0,0)*{\bullet}
,(0,0.5)*{a}
,(-4.5,3.5)*{\text{mirror}}
,(4.5,-3.5)*{\text{mirror}}
,(0.75,2)*{\text{splitter}}
,(-0.75,-2)*{\text{splitter}}
\ar (-3,3);(0,3)
\ar@{-} (0,3);(2.5,3)
\ar (3,-3);(3,0)
\ar@{-} (3,0);(3,2.5)
\ar (-2.5,-3);(0,-3)
\ar@{-} (0,-3);(3,-3)
\ar (-3,-2.5);(-3,0)
\ar@{-} (-3,0);(-3,3)
\ar@{-} (2,2.5);(3.5,4)
\ar@{-} (2.5,2);(4,3.5)
\ar@{-} (2,2.5);(2.5,2)
\ar@{-} (3.5,4);(4,3.5)
\ar (3.5,3);(6,3)
\ar@{-} (7,4);(8,3)
\ar@{-} (7,2);(8,3)
\ar@{-}@/_6pt/ (7.25,3.75);(7.25,2.25)
\ar@{-}@/^3pt/ (6.925,3.33);(6.925,2.625)
\ar@{-} (-2,-2.5);(-3.5,-4)
\ar@{-} (-2.5,-2);(-4,-3.5)
\ar@{-} (-2,-2.5);(-2.5,-2)
\ar@{-} (-3.5,-4);(-4,-3.5)
\ar (-6,-3);(-3.5,-3)^{bbbb\cdots}
\ar@{-} (-4,2);(-2,4)
\ar@{-} (4,-2);(2,-4)
\end{xy}\]
\caption{Mach-Zehnder}
\end{subfigure}
\newline
\begin{subfigure}{\textwidth}
\[\begin{xy}<5mm,0mm>:
(1.5,0)*{\sbendh}
,(2,0.5)*{\zbendh}
,(2.75,0)*{\sbendh}
,(3.25,0.5)*{\zbendh}
,(1.75,4)*\xycircle<7pt>{}
,(1.75,2)*{\text{detector}}
,(1.75,-0.5)*{b\text{-edge mode}}
,(9.625,2.375)*{b\text{-edge mode}}
,(1.5,1.5)*{\zbendh}
,(2,1)*{\sbendh}
,(2.75,1.5)*{\zbendh}
,(3.25,1)*{\sbendh}
,(5.25,1.5)*{\bullet}
,(5.25,2)*{a}
\ar (0.5,0);(3,0)
\ar@{-} (5,0);(5.5,0)
\ar@{-} (7.5,0);(10,0)
\ar@{-} (2.5,3);(3,3)
\ar@{-} (5,3);(5.5,3)
\ar (10,3);(7.5,3)
\ar@{-} (1,3);(0.5,3)
\ar@{-} (2.5,3.5);(2.5,2.5)
\ar@{-} (1,3.5);(2.5,3.5)
\ar@{-} (1,2.5);(2.5,2.5)
\ar@{-} (1,3.5);(1,2.5)
\ar (1.75,4);(2,4.4)
\ar_{t_1} (4,1.25);(4,1.75)
\ar_{t_2} (6.5,1.25);(6.5,1.75)
\end{xy}\]
\caption{Fabry-Perot}
\end{subfigure}
\caption{}
\label{fig:1}
\end{figure}

In contrast, a projective measurement $\operatorname{Proj}_a$ to particle type $a$ is the Hermitian orthogonal projection to that particle sector.  It is visualized as occurring by bringing some external prob, such as an STM tip, up to an isolated point-like anyon and directly detecting some, perhaps non-universal, signature of that particular particle type in that particular system.  For example, even for an electrically neutral $\psi$, in $\nu = \frac{5}{2}$ fractional quantum Hall effect (FQHE), higher moments of the electric field might provide a signature.  In any case, it is this hope which has led to the projective measurement model.

We now show that within the UMTC formalism, any protocol using projective measurement can be efficiently simulated by a protocol which instead uses interferometric measurement.

To do this we first need to define, through a density matrix diagram, the asymptotic topological action of $\operatorname{Int}_a$.  The diagrams, Figure \ref{fig:2b} and on, have a $|\text{ket}\rangle\langle\text{bra}|$ aspect when read from top to bottom. The diagrams in $|\text{ket}\rangle\langle\text{bra}|$ format, of course, represent operators (density matrices), and Figure \ref{fig:2a} represents a state vector ($|\text{ket}\rangle$).  It should be noted that such representations of operators and states obey topological rules \cite{W} and may or may not correspond bit by bit to a physical process.

We start with a vacuum and create a, perhaps complex, system of anyons which we divide into two halves ``inside'' and ``outside'' the interferometer.  In Figure \ref{fig:2} these two halves at any given time are depicted simply as points, but they may represent composite anyons, in dual groups, drawn out of the vacuum.

\begin{figure}[hbpt]
\begin{subfigure}{0.4\textwidth}
\[\begin{xy}<5mm,0mm>:
(0,0)*{\vloop[-3]}
,(0,0)*{\bullet}
,(3,0)*{\bullet}
,(0,0.75)*{\text{dual}}
,(3,0.625)*{\text{anyon}}
,(3.75,1.5)*{\text{(composite)}}
,(5.5,-3.5)*{\text{vacuum}}
\ar_{\text{time}} (5.5,-2.5);(5.5,0)
\end{xy}\]
\caption{$|\text{ket}\rangle$}
\label{fig:2a}
\end{subfigure}
\begin{subfigure}{0.4\textwidth}
\[\begin{xy}<5mm,0mm>:
(0,0)*{\vloop[-2]}
,(0,0)*{\bullet}
,(2,0)*{\bullet}
,(0,-5.5)*{\bullet}
,(2,-5.5)*{\bullet}
,(0,-2.75)*{\vloop[2]}
,(3.5,-1)*{|\text{ket}\rangle}
,(3.5,-4.5)*{|\text{bra}\rangle}
\end{xy}\]
\caption{$\rho=|\text{ket}\rangle\langle\text{bra}|$}
\label{fig:2b}
\end{subfigure}
\caption{}
\label{fig:2}
\end{figure}

From \cite{B1,B2}, $\operatorname{Int}_a$ asymptotically transforms from Figure \ref{fig:2b} to Figure \ref{fig:3}, assuming that the measurement outcome $a$, indeed, has nonzero probability.  In Figure \ref{fig:3} and below, we drop overall nonzero scalars from the diagrams.

\begin{figure}[hbpt]
\begin{subfigure}{0.15\textwidth}
\[\begin{xy}<5mm,0mm>:
(0,0)*{\vcap[2]|(0.2)\khole}
,(0,-0.5)*{\vcap[-2]|(0.8)\khole}
,(0,1.25)*{\vloop[-2]|(0.2)\khole|(0.91)\khole}
,(0,-1.75)*{\vloop[2]|(0.025)\khole|(0.8)\khole}
,(0.8125,0.9325)*{\vcap[0.75]|(0.5)\khole}
,(0.8125,0.9325)*{\vcap[-0.75]}
,(0.8125,-1.4325)*{\vcap[0.75]|(0.5)\khole}
,(0.8125,-1.4325)*{\vcap[-0.75]}
,(3.125,1.865)*{\omega_a}
,(3.125,-2.865)*{\omega_a}
,(2.75,-0.5)*{\omega_0}
\ar@{-} (0,0);(0,-1)
\ar@{-} (2,0);(2,-1)
\end{xy}\]
\caption{}
\end{subfigure} $= \Sigma_k F^{\bar{a},{a}}_{\bar{a},{a};0,k}$
\begin{subfigure}{0.1\textwidth}
\[\begin{xy}<5mm,0mm>:
(0,-0.5)*{\bar{a}}
,(2,-0.5)*{a}
,(0,4.5)*{\bar{a}}
,(2,4.5)*{a}
,(0.625,1.1875)*{\hcap[0.75]}
,(0.625,1.1875)*{\hcap[-0.75]|(0.5)\khole}
,(1.25,1.125)*{\omega_0}
\ar@{-} (0,0);(0,4)
\ar@{-} (2,0);(2,4)
\ar@{-}|(0.75)\hole^(0.23){k} (0,2);(2,2)
\end{xy}\]
\caption{}
\end{subfigure} $\underset{\text{dropping constants}}{=}$
\begin{subfigure}{0.15\textwidth}
\[\begin{xy}<5mm,0mm>:
(0,-0.5)*{\bar{a}}
,(2,-0.5)*{a}
,(0,4.5)*{\bar{a}}
,(2,4.5)*{a}
\ar@{-} (0,0);(0,4)
\ar@{-} (2,0);(2,4)
\end{xy}\]
\caption{}
\label{fig:3c}
\end{subfigure}
\caption{}
\label{fig:3}
\end{figure}

Notation: $\omega_a$ is the $a^\text{th}$ row of the normalized $S$-matrix which operates as a projector onto the $a$ particle:
\[\begin{xy}<5mm,0mm>:
(-0.25,0)*{\vcap|(0.5)\khole}
,(-0.25,0)*{\vcap-}
,(0,1.5)*{b}
\ar@{-}|(0.25)\hole (0,-1);(0,1)
\end{xy}
\;\;\;\omega_a = \delta_{a,b}\;
\begin{xy}<5mm,0mm>:
(0,1.5)*{b}
\ar@{-} (0,-1);(0,1)
\end{xy}.\]

The presence of $\omega_a$ is expected---it projects onto the $a$-particle type sector.  The $\omega_0$ loop has long been regarded as an unfortunate but unavoidable consequence of running the interferometer.  $\omega_0$ encodes a kind of decoherence between inside and outside caused by the intervening stream of prob particles $b$.  In \cite{B2}, this severing of charge lines $k$ running from the inside to the outside of the interferometer was called \emph{anyonic charge line decoherence}.  Topologically, $\omega_0$ surgers the $a$-lines (Figure \ref{fig:3c}) so that $a$ and $\bar{a}$ have forgotten that they came out of the vacuum together---they are no longer correlated.

The presence of this $\omega_0$-loop leads to a general supposition in the community that $\operatorname{Int}_a$ and $\operatorname{Proj}_a$ were incomparable: $\operatorname{Int}_a$ permits non-demolition measurement, as the internal correlations within the two groups of quasiparticles, inside and outside, are \emph{not} disturbed, whereas in order to obtain a localized particle on which to apply $\operatorname{Proj}_a$, the internal structure of the quasiparticle group to be measured would first need to be destroyed by a series of fusions (which produces decoherence even if the fusion outcomes are presumed not to be observed).  In contrast, $\operatorname{Proj}_a$ does \emph{not} cause anyonic charge line decoherence between the measured subsystem from its complement.  Each seemed to have its own peculiar advantages and disadvantages with respect to the preservation of quantum information.

However, we will now show that the anyonic charge line decoherence of $a$ and $\bar{a}$ can be reversed (oddly, this oximoron \emph{is} possible in a topological system) by additional interferometric measurements.  The key observation is that if after $\operatorname{Int}_a$ we interferometrically measure the collective state of $\bar{a}\cup a$, there is still a nonzero probability of observing the outcome $0$, the trivial particle.  This is because if the two lines in Figure \ref{fig:3c} are recoupled, the $F$-symbol, $F^{\bar{a},a}_{\bar{a},{a};0,0} \neq 0$, reflecting the fact that a particle and its uncorrelated antiparticle may fuse into the vacuum.  (This, in fact, is the definition of an antiparticle.)  Suppose we measure $\bar{a}\cup a$ and observe $0$, so that $\operatorname{Int}_0$ is applied to the system.  The result is given in Figure \ref{fig:4}.

\begin{figure}[hbpt]
\[\begin{xy}<5mm,0mm>:
(0.5,0.25)*{\hloop-|(0.15)\khole}
,(0.5,0.25)*{\hloop|(0.85)\khole}
,(3,0)*{\omega_0}
,(0.25,-2.5)*{\bar{a}}
,(1.75,-2.5)*{a}
,(0.25,2.5)*{\bar{a}}
,(1.75,2.5)*{a}
\ar@{-}|(0.375)\hole (0.25,-2);(0.25,2)
\ar@{-}|(0.375)\hole (1.75,-2);(1.75,2)
\end{xy}
= \Sigma F^{\bar{a},{a}}_{\bar{a},{a};0,k}
\begin{xy}<5mm,0mm>:
(2.5,0)*{\omega_0}
,(0,1)*{\vcap[-2]}
,(0,-1)*{\vcap[2]}
,(0.4,0.125)*{\hloop[-0.5]}
,(0.6,0.125)*{\hloop[0.5]}
,(0,2.5)*{\bar{a}}
,(2,2.5)*{a}
,(0,-2.5)*{\bar{a}}
,(2,-2.5)*{a}
\ar@{-}_(0.8125)k|(0.375)\hole (1,-1);(1,1)
\ar@{-} (0.8,-0.25);(1.2,-0.25)
\end{xy}
= F^{\bar{a},{a}}_{\bar{a},{a};0,0}
\begin{xy}<5mm,0mm>:
,(0,1)*{\vcap[-2]}
,(0,-1)*{\vcap[2]}
,(0,2.5)*{\bar{a}}
,(2,2.5)*{a}
,(0,-2.5)*{\bar{a}}
,(2,-2.5)*{a}
\end{xy}
\underset{\text{dropping constants}}{=}
\begin{xy}<5mm,0mm>:
,(0,1)*{\vcap[-2]}
,(0,-1)*{\vcap[2]}
,(0,2.5)*{\bar{a}}
,(2,2.5)*{a}
,(0,-2.5)*{\bar{a}}
,(2,-2.5)*{a}
\end{xy}\]
\caption{}
\label{fig:4}
\end{figure}

We see that $\operatorname{Int}_0$ has restored us to the situation we would have been in if the $\omega_0$ loop associated to the initial $\operatorname{Int}_a$ had \emph{not} been present.  The decoherence  has been reversed.

Of course it is not certain that the second measurement will result in $\operatorname{Int}_0$; other outcomes $\operatorname{Int}_k$ are also possible.  But now there is an easy ``ping-pong'' strategy (referred to as ``forced measurement'' in \cite{B4}): Bounce back and forth between measuring the initial inside---which will always return outcome $a$, $\operatorname{Int}_a$ will be applied on the odd steps of this cycle, and measuring the entire system $\bar{a}\cup a$.  Each of the odd steps decoheres the inside and outside of the initial interferometer and returns the density matrix to that shown in Figure \ref{fig:3c}.  Actually, since on the odd steps $3, 5, 7,\ldots$, there is no doubt that topological charge $a$ will be measured, it is \emph{not} necessary to \emph{read} the output of the interferometer, but merely to run the probe particles around the initial loop, thus producing anyonic charge line decoherence.  Each of these even step measurements on the entire system constitute an \emph{independent} chance to apply $\operatorname{Int}_0$.  Because of independence, the tail event that after $2t$ measurements $\operatorname{Int}_0$ has not been applied decays exponentially in $t$.  The exact exponential rate is easily calculated from the data of any particular UMTC.  Since $F^{\bar{a},a}_{\bar{a},{a};0,0} = \frac{1}{d_a}$, charge $0$ is observed at each step $2t$ with probability $p = \left(\frac{1}{d_a}\right)^2$, so the exponential rate of decay in $t$ is $2\log_e\left(1 - \frac{1}{d_a^2}\right)$.

\section{Conclusion and outlook}

We have shown how to ``projectively''\footnote{We place ``projectively'' in quotes because unlike the usual usage in the arena of anyonic systems, this measurement is both nonlocal and nondemolitional.  It is projective in the usual quantum mechanical sense of Hermitian orthogonal projection onto an eigenbasis, in this case the eigenbasis of total topological charge.} measure the total topological charge with repeated interferometric measurements of groups of anyons without decohering the group from its complement.  In contrast, projective measurement of anyonic charge is in the usual model \cite{K} limited to projecting to the topological charge of a single anyon.  A priori, this looks like a strictly \emph{weaker} operation.

It should be remarked, though, that as with all issues of complexity, at this point in history there are no ``lower bounds.''

Any proof that it is impossible to simulate interferometric measurement by projective measurement would necessarily rely on complexity assumptions.  We regard this as an area for future work.

However, evidence of the enhanced strength of interferometric measurement is presented in a series of papers \cite{B5,L} on universal gate systems for qubit and qutrit systems within $SU(2)_4$ and its Jones-Kauffman partner.  Previously a universal protocol for a certain qutrit within $SU(2)_4$ was found \cite{CW} using projective measurement but the argument appears quite special and not applicable to qubits.

We call attention to a shortcut for graphically exploring interferometry protocols.  Because of the iterative process we have just described for eliminating the decohering $\omega_0$-loops associated to interferometric measurement, one may proceed---in the manner of a person writing computer code in a higher order language---only to manipulate the $|\text{ket}\rangle$ which describes the current state of the system of anyons.  The $|\text{ket}\rangle$ is used at any given time to describe the state that has been pulled out of the vacuum.  It is not necessary to double the diagram by adding the dual bra (and the linking $|\text{ket}\rangle\langle\text{bra}|$ by $\omega_0$-loops).  Any $\omega_0$-loop will eventually be removed by some even numbered step of our protocol.  Thus it is not really necessary to draw the $\omega_0$-loops, or even the $\langle\text{bra}|$, but merely to keep track of the $|\text{ket}\rangle$.  In the end, if a density matrix $\rho$ is desired, one may obtain $\rho$ as the outer product of the final $|\text{ket}\rangle_\text{final}$ with its dual $\langle\text{bra}|_\text{final}$, $\rho = |\text{ket}\rangle_\text{final}\langle\text{bra}|_\text{final}$.

\bibliographystyle{acm}
\bibliography{Interferometric_versus_Projective_Measurement_of_Anyons.bbl}

\end{document}